\documentclass[final,10pt,a4paper]{elsart}

\usepackage{amsmath,amssymb} 

\newcommand{\pd}[3]{ \frac{ \partial^{#3}{#1} }{ \partial {#2}^{#3} } }

\begin{document}

\begin{frontmatter}

\title{Theoretical Criteria for the Occurrence of Turbulence in Burger's Equation}
\author[UP]{J. C. Imperio}
\author[UP]{Mikhail P. Solon}
\author[UP]{A. Laganapan}
\author[UP]{J. P. H. Esguerra}
\author[DTS]{A. Muriel}
\address[UP]{Theoretical Physics Group, National Institute of Physics\\ University of the Philippines Diliman, Quezon City 1101}
\address[DTS]{Data Transport Systems\\347 East 62nd Street, New York, NY 10021, USA}

\begin{abstract}
Throughout the history of the study of turbulence in fluid dynamics, there has yet to arise a unique definition or theoretical criterion for this important phenomenon. There have been interesting conjectures made by Ruelle \cite{Ruelle}, Muriel \cite{QKMMuriel}, and Getreuer, Albano and Muriel \cite{GAM}, however, attempting to provide the sufficient criteria for the onset of turbulence. In this paper, a classic equation in fluid dynamics, Burger's equation, is solved in one and two dimensions, and these conjectures are illustrated. This illustration supports these conjectures by showing that the proposed criteria do arise mathematically from the solutions of an equation modelling fluid flows.
\end{abstract}
\end{frontmatter}

To this day, there is no unique, universally acceptable theoretical criterion for the occurrence of turbulence \cite{WT}. However, Ruelle \cite{Ruelle} conjectured that when the time derivative of the velocity inside a fluid becomes infinite, the system becomes turbulent. A less stringent criterion was proposed by Muriel \cite{QKMMuriel}, which is that if the velocity reverses (even slowly), turbulence could result. The latter criterion seems to have been observed in pipe experiments \cite{RadMuriel, EqMuriel}. Recently, an intriguing suggestion has been made by Getreuer, Albano and Muriel \cite{GAM} that a multiple-valued velocity field may represent turbulence. In this Letter, we explore each of these conjectures using the example of Burger's equation. 

Burger's equation is a fundamental PDE from fluid mechanics. It can be obtained from the more general Navier-Stokes equation under the proper set of assumptions. It has been extensively studied over the years, both generally (as in \cite{Burgers1}), and in particular with respect to turbulence \cite{Burgers2, Burgers3}.

In this paper, the stationary states of Burger's equation in one and two dimensions are studied.  Upon application of periodic boundary conditions, the solutions are shown to become multi-valued, which allows for the possibility of velocity reversal. We also obtain the time-dependent solutions, and show infinite time derivatives are possible, with the proper choice of arbitrary solution constants. 

The development of the above results follows.

The Navier-Stokes equation can be written as follows:

\begin{align}
\rho\left(\pd{\bf{v}}{t}{}+ \left(\bf{v}\cdot\nabla\right)\bf{v}\right)=-\nabla p + \nabla \cdot \mathbb{T} + \bf{f}.
\end{align}

Here $\rho$ represents the mass density of the fluid, $\bf{v}$ is the velocity, $p$ is the pressure, $\mathbb{T}$ is the stress tensor, and $\bf{f}$ represents the sum of any other forces.

We make the following simplifying assumptions:

\begin{align}
\nabla \cdot \mathbb{T} & = \nabla^2 \bf{v}, \label{eq:inc}\\
\bf{f} & = 0, \label{eq:fz} \\
\nabla p & = 0. \label{eq:prc}
\end{align}

In other words, we consider a Newtonian, incompressible fluid (\ref{eq:inc}) under no external forces (\ref{eq:fz}) or pressure gradient (\ref{eq:prc}). 

Setting the velocity constant in time

\begin{align}
\pd{\bf{v}}{t}{}=0,
\end{align}

\noindent is equivalent to solving for the stationary states of the system. In Cartesian coordinates and considering the 1-D case, we get

\begin{align}
\rho \left(v \pd{v}{x}{} \right) = \mu \pd{v}{x}{2},
\end{align}

\noindent a solution of which is given by

\begin{align}
v(x) = \sqrt{\frac{2 C_1 \mu}{\rho}} \tan \left( \sqrt{\frac{C_1 \rho}{2 \mu}} \left( x + C_2 \right) \right). \label{eq:1Dsol}
\end{align}

Imposing periodic boundary conditions

\begin{align}
v(x) = v(x+L)
\end{align}

\noindent yields the value of the constant $C_1$:

\begin{align}
C_1 = \frac{2 \mu \pi^2 k^2}{\rho L^2},
\end{align}

\noindent where k is any nonzero integer. Upon substitution, solution (\ref{eq:1Dsol}) becomes

\begin{align}
v_{k}(x) = \frac{2 \mu k \pi}{\rho L} \tan \left( \frac{k \pi x}{L} \right). \label{eq:1Dsolbc}
\end{align}

Here, $C_2$ is dropped, equivalent to setting $v(0) = 0$. 

Note that we now have a family of solutions indexed by the integer $k$. This means that the velocity field can now be considered a multi-valued function.

Considering the same problem in two dimensions, we have the following set of equations for the stationary states:

\begin{align}
\rho \left( u_x \pd{u_x}{x}{} + u_y \pd{u_x}{y}{} \right) = \mu\left( \pd{u_x}{x}{2} + \pd{u_x}{y}{2} \right), \\
\rho \left( u_x \pd{u_y}{x}{} + u_y \pd{u_y}{y}{} \right) = \mu\left( \pd{u_y}{x}{2} + \pd{u_y}{y}{2} \right),
\end{align}

\noindent where the total velocity function is given by $\bf{u} = u_{x} \hat{i} + u_{y} \hat{j}$. Solutions to this system of differential equations are given by

\begin{align}
u_x(x,y) & =C_4 + C_5 \tanh\left( C_1 + C_2 x + C_3 y \right), \\
u_y(x,y) & =-\frac{C_2 C_4}{C_3} - \frac{\rho C_2 C_5 + 2 \mu \left(C_2^2 + C_3^2 \right)}{\rho C_3} \tanh(C_1 + C_2 x + C_3 y).
\end{align}

Setting the arbitrary constants $C_1, C_2$ and $C_3$ to be imaginary, (i.e., $C_j=iK_j$, where $K_j$ is real) leads to the hyperbolic tangent becoming a tangent function:

\begin{align}
u_x(x,y) & =C_4 + C_5 \tan\left( K_1 + K_2 x + K_3 y \right), \\
u_y(x,y) & =-\frac{K_2 C_4}{K_3} - \frac{\rho K_2 C_5 - 2 \mu \left(K_2^2 + K_3^2 \right)}{\rho K_3} \tanh(K_1 + K_2 x + K_3 y).
\end{align}

Boundary conditions along both the x and y directions can now be imposed, similar to what was done in the 1D case, to obtain

\begin{align}
u_x(x,y) & =K_4 + K_5 \tan\left( K_1 + \frac{k_x \pi}{L_x} x + \frac{k_y \pi}{L_y} y \right), \\
u_y(x,y) & = - \frac{L_y \left( \rho \frac{k_x K_5} L_x + 2 \mu \pi \left((\frac{k_x}{L_x})^2 + (\frac{k_y}{L_y})^2 \right) \right)}{\rho k_y} \tan(K_1 + \frac{k_x \pi}{L_x} x + \frac{k_y \pi}{L_y} y) \notag \\
& \quad -\frac{k_x L_y K_4}{L_x k_y}.
\end{align}

These 2D solutions are now indexed by two integers, $k_x$ and $k_y$, and can again be considered multi-valued.

These 1D and 2D solutions may then be interpreted to be possible states of the system, with the possibility of transitions between them, following Getreuer, Albano and Muriel \cite{GAM}. This interpretation then allows for an explanation of velocity reversal (a signature of turbulence) - the system simply jumps from a positive solution to a negative one.

Finally, in comparison to our results, there is yet another multi-valued character in one of the branches of the GAM solution. In one branch, which corresponds to our integer labels, a bifurcation of the solution is observed. Such a bifurcation is a characteristic signature of chaos, but which we do not find in our example. In this example, the multi-valued nature of the solutions is not an intrinsic characteristic, but rather due to the imposition of periodic boundary conditions.

The time-dependent 1-D Burger's equation takes the form

\begin{align}
\rho \left(\pd{v}{t}{} + v \pd{v}{x}{} \right) = \mu \pd{v}{x}{2},
\end{align}

\noindent with the corresponding solution

\begin{align}
v(x,t)=-\frac{C_3}{C_2}-\frac{2 \mu C_2}{\rho}\tanh(C_1+C_2x+C_3t). \label{eq:1Dtdsol}
\end{align}

Note that with the choice of parameters $C_1=0, C_2=ik\pi/L,$ and $C_3=0$, this time-dependent solution (\ref{eq:1Dtdsol}) reduces to the previously obtained result (\ref{eq:1Dsolbc}). In addition, by making the arbitrary constants $C_1$, $C_2$, and $C_3$ imaginary, as before, we obtain an expression involving the tangent instead of the hyperbolic tangent:

\begin{align}
v(x,t)=-\frac{K_3}{K_2}-\frac{2 \mu K_2}{\rho}\tan(K_1+K_2x+K_3t).
\end{align}

The time derivative of this solution is

\begin{align}
\pd{v}{t}{}=-\frac{2 \mu K_2 K_3}{\rho}\left( 1 + \tan^2(K_1+K_2x+K_3t) \right),
\end{align}

\noindent which evidently approaches infinity for particular values of x and t. 

In 2D, the time-dependent Burger's equations are

\begin{align}
\rho \left( \pd{u_x}{t}{} + u_x \pd{u_x}{x}{} + u_y \pd{u_x}{y}{} \right) = \mu\left( \pd{u_x}{x}{2} + \pd{u_x}{y}{2} \right), \\
\rho \left( \pd{u_y}{t}{} + u_x \pd{u_y}{x}{} + u_y \pd{u_y}{y}{} \right) = \mu\left( \pd{u_y}{x}{2} + \pd{u_y}{y}{2} \right),
\end{align}

\noindent with the corresponding solutions

\begin{align}
u_x(x,y) & = - \frac{\rho C_3 C_6 + 2 \mu \left( C_2^2 + C_3^2 \right)}{\rho C_2} \tanh \left( C_1 + C_2 x + C_3 y + C_4 t \right) \notag \\
& \quad -\frac{C_4 + C_3 C_5}{C_2}, \\
u_y(x,y) & = C_5 + C_6 \tanh \left( C_1 + C_2 x + C_3 y + C_4 t \right).
\end{align}

We may again choose imaginary constants within the argument of the hyperbolic tangent to transform the function into the tangent. The time derivatives of the transformed solutions are then

\begin{align}
\pd{u_x}{t}{} & = - \frac{K_4 \left( \rho K_3 K_6 + 2 \mu \left( K_2^2 + K_3^2 \right) \right)}{\rho K_2} \left(1 + \tan^2 \left( K_1 + K_2 x + K_3 y + K_4 t \right) \right), \\
\pd{u_y}{t}{} & = K_4 K_6 \left( 1 + \tan^2 \left( K_1 + K_2 x + K_3 y + K_4 t \right) \right).
\end{align}

These time derivatives again clearly become infinite at certain values of $x$, $y$, and $t$. 

The points at which the time derivatives of these 1-D and 2-D solutions become infinite may very well be points where the onset of turbulence occurs, as hypothesized by Ruelle \cite{Ruelle}.

In conclusion, we exhibit an example of the Ruelle conjecture, illustrate the modification proposed by Muriel, and quite interestingly, satisfy the conjecture of Getreuer, Albano and Muriel that multi-valued velocity fields could possibly describe turbulence.

\end{document}